\documentclass[sigconf,10pt,nonacm,screen]{acmart}
\renewcommand\footnotetextcopyrightpermission[1]{}

\usepackage{microtype}
\usepackage[htt]{hyphenat}
\usepackage{xspace}
\usepackage{makecell}
\usepackage{threeparttable}
\usepackage{listings}
\usepackage{xcolor}
\usepackage{caption}
\usepackage{subcaption}
\usepackage{placeins}
\usepackage{float}
\usepackage{enumitem}
\usepackage{needspace}
\usepackage{times}  
\usepackage{hyperref}
\usepackage{latexsym}
\usepackage[export]{adjustbox}
\usepackage{multirow}
\usepackage{tikz}
\usepackage{array}
\usepackage{comment}
\usepackage{url}
\usepackage{xurl}
\usepackage[normalem]{ulem}
\usepackage{mdwlist}
\usepackage{algorithm}
\usepackage{booktabs}
\usepackage[noend]{algpseudocode}
\usepackage{centernot}
\usepackage{fancyvrb}
\usepackage{alltt}
\usepackage[ampersand]{easylist}
\usepackage{fancybox}
\usepackage{graphicx}
\usepackage{tcolorbox}
\usepackage{balance}
\usepackage{xcolor}
\usepackage{verbatim}
\usepackage{breakurl}
\usepackage{xcolor}
\usepackage{listings}
\usepackage{framed}
\usepackage{lipsum} 
\usepackage[framemethod=tikz]{mdframed}
\usepackage{tabularx}

\definecolor{codeblue}{rgb}{0,0,1}
\definecolor{codegreen}{rgb}{0,0.6,0}
\definecolor{codegray}{rgb}{0.5,0.5,0.5}
\definecolor{codepurple}{rgb}{0.58,0,0.82}
\definecolor{backcolour}{rgb}{0.95,0.95,0.92}
\definecolor{nocolor}{rgb}{1,1,1}

\definecolor{red}{rgb}{0.6,0,0} 
\definecolor{blue}{rgb}{0,0,0.6}
\definecolor{green}{rgb}{0,0.8,0}
\definecolor{cyan}{rgb}{0.0,0.6,0.6}
\definecolor{lightgray}{gray}{0.98}
\definecolor{lightblue}{rgb}{0.13, 0.67, 0.8}
\definecolor{lightorange}{RGB}{255,247,230}
\definecolor{codegreen}{rgb}{0,0.6,0}
\definecolor{codegray}{rgb}{0.5,0.5,0.5}
\definecolor{codepurple}{rgb}{0.58,0,0.82}
\definecolor{keywordcolor}{RGB}{94,20,64}
\definecolor{bluekeywords}{rgb}{0,0,1}
\definecolor{greencomments}{rgb}{0,0.5,0}
\definecolor{redstrings}{rgb}{0.64,0.08,0.08}
\definecolor{xmlcomments}{rgb}{0.5,0.5,0.5}
\definecolor{types}{rgb}{0.17,0.57,0.68}
\definecolor{KWColor}{RGB}{0,0,255}
\definecolor{AnnotationColor}{RGB}{0,137,180}
\definecolor{BlackColor}{RGB}{0,0,0}
\definecolor{CommentColor}{rgb}{0.12,0.38,0.18}
\definecolor{StringColor}{rgb}{0.06,0.10,0.98}
\definecolor{darkred}{rgb}{0.65,0,0}
\definecolor{lightgrey}{rgb}{0.8,0.8,0.8}
\definecolor{marmalade}{RGB}{193,101,18}
\definecolor{peach}{RGB}{250,217,193}
\definecolor{lime}{RGB}{220,237,193}

\newenvironment{packeditemize}{
\begin{itemize}[leftmargin=1em]
  \setlength{\itemsep}{2pt}
  \setlength{\parskip}{2pt}
  \setlength{\parsep}{2pt}
  \setlength{\topsep}{2pt}
}{\end{itemize}}

\lstdefinestyle{P4}{
  showspaces=false,
  showtabs=false,
  tabsize=2,
  columns=flexible,
  keepspaces=true,
  language={Python},
  numbers=left,
  basicstyle=\ttfamily\scriptsize,
  commentstyle=\itshape\color{gray}\ttfamily\scriptsize,  
  stringstyle=\color{codepurple},                          
  showstringspaces=false,
  upquote=true,
  xleftmargin=1.2em,
  framexleftmargin=1.5em,
  numberstyle=\scriptsize\color{gray},
  keywords={     StateMachine, class, return},
  keywords=[2]{  bool, str, int, List, SM, self},
  keywords=[3]{  States, Transitions, def},
  keywords=[4]{  read, assert, call, write},
  keywordstyle=\color{BlackColor}\bfseries,
  keywordstyle=[2]\color{codeblue},
  keywordstyle=[3]\color{red}\bfseries,                   
  keywordstyle=[4]\color{codegreen},
  escapeinside={/*@}{@*/},
}

\mdfdefinestyle{background}{backgroundcolor=lightorange,innerrightmargin=0cm,innertopmargin=-0.1cm,innerbottommargin=-0.10cm,leftmargin=+0cm, roundcorner=2pt}

\usepackage{enumitem}
\usepackage[framemethod=tikz]{mdframed}

\usepackage[textsize=tiny,textwidth=0.6in]{todonotes}
\newcommand{\archit}[1]
{\todo[color=teal!20]{Archit: #1}}
\newcommand{\yiming}[1]
{\todo[color=orange!50]{Yiming: #1}}
\newcommand{\zy}[1]{\todo[color=pink!50]{Zhenning: #1}}
\newcommand{\sr}[1]{\todo[color=violet!20]{Sylvia: #1}}
\newcommand{\ang}[1]{\todo[color=blue!20]{Ang: #1}}
\newcommand{\sarah}[1]{\todo[color=green!20]{Sarah: #1}}

\newcommand{\sys}{\textsc{CloudEmu}\xspace}

\makeatletter
\renewcommand\@formatdoi[1]{\ignorespaces}
\makeatother

\makeatletter
\def\@copyrightspace{\relax}
\makeatother

\AtBeginDocument{%
  }

\newcounter{defn}[section]
\renewcommand{\thedefn}{\arabic{section}.\arabic{defn}}
\newenvironment{defn}[2][]{%
\refstepcounter{defn}%
\ifstrempty{#1}%
{\mdfsetup{%
frametitle={%
\tikz[baseline=(current bounding box.east),outer sep=0pt]
\node[anchor=east,rectangle,fill=purple!20]
{\strut Definition~\thedefn};}}
}%
{\mdfsetup{%
frametitle={%
\tikz[baseline=(current bounding box.east),outer sep=0pt]
\node[anchor=east,rectangle,fill=purple!20]
{\strut #1};}}%
}%
\mdfsetup{innertopmargin=10pt,linecolor=purple!20,%
linewidth=2pt,topline=true,%
frametitleaboveskip=\dimexpr-\ht\strutbox\relax
}
\begin{mdframed}[]\relax%
\label{#2}}{\end{mdframed}}

\title{Automated Synthesis of Cloud Emulators}  

\author{Archit Bhatnagar\hspace{2mm} Zhenning Yang\hspace{2mm} Sarah McClure$^\ddag$ \hspace{2mm} \\  Yiming Qiu$^\dag$  \hspace{2mm} Sylvia Ratnasamy$^\ddag$ \hspace{2mm}  Ang Chen \vspace{2mm}}

\affiliation{
\textit{{University of Michigan\hspace{2mm}
$^\dag$The University of Hong Kong \hspace{2mm} $^\ddag$University of California, Berkeley}}\country{\unskip}
}

\begin{document}

\begin{abstract} 
DevOps programming (e.g., using CLI/API scripts or IaC frameworks) is key to cloud infrastructure management.  
Unlike traditional programming tasks, DevOps program testing needs provisioning and execution against actual cloud resources, which is often time-consuming, unsafe, and costly. 
Cloud emulators have gained popularity for easing DevOps program testing; they are generally API-level mocks that can execute DevOps programs in a local environment. 
Still, building these emulators remains challenging: developers must manually interpret extensive cloud documentation and handcraft logic for each service, API, and their interaction. This does not scale to the complexity of the cloud, which is further a moving target as the services and APIs evolve. 
\sys is an automated approach that constructs emulators based on cloud documentation via neurosymbolic code synthesis. 
The key idea is to combine LLMs' general strengths in documentation understanding and code generation with cloud-specific symbolic abstractions that suppress hallucinations and enforce precision at scale, while using the real cloud as an oracle for automated testing, repair, and alignment. 
Our evaluation shows the effectiveness of \sys on major cloud provider (AWS and GCP) services in both coverage and accuracy. \sys outperforms the existing leading tool LocalStack, which was manually developed by a large team of engineers over a decade. 


\end{abstract}

\maketitle
\pagestyle{plain}

\section{Introduction}
\label{sec:intro}

Building and maintaining cloud infrastructure (e.g., virtual machines, gateways) is an essential task. DevOps engineers perform these tasks programmatically---e.g., using CLI commands or Python scripts that directly invoke cloud-level APIs~\cite{terra-api, cloud_control} exposed by the provider, or using Infrastructure-as-Code (IaC) frameworks~\cite{pulumi, terraform, cloudformation} (e.g., Terraform) that compile infrastructure configurations to API calls.
However, testing against the cloud is expensive~\cite{cloud_test_expensive}, and resource provisioning can be time-consuming. Prices and provisioning time can further increase for resources in high demand~\cite{skypilot}. 
This complicates DevOps programming and reduces velocity.

To enable no-risk, no-cost, and high-velocity cloud development, \textit{cloud emulators}~\cite{localstack, moto,azurite} are quickly 
gaining traction. Emulators mimic the cloud by exposing identical API interfaces to DevOps programs and simulating their execution in a mock environment, providing a lightweight backend without going through the real cloud. In order to emulate a resource (e.g., VM), emulator developers sift through cloud documentation, identify target APIs, and handcraft the mockup logic based on their understanding of the expected behavior.
Interdependent resources (e.g., VM is associated with Subnet and VPC) further need to be emulated in relation to each other. 

While this is a laudable effort, existing practices of emulator development cannot catch up to the complexity and dynamicity of the cloud ecosystem. For instance,  
AWS alone provides 240 services~\cite{aws_services}, and a service can expose up to 200 APIs; Azure and GCP exhibit a similar level of complexity. 
Market competition means that providers often add new services and upgrade existing ones, making the cloud a moving target~\cite{cloud_evolve}. 
Each cloud provider also features a different set of 
services and APIs, and more players are entering the 
cloud market. 
To ensure reliability at scale, tenants often construct multi-cloud deployments, which further propagates the need for DevOps testing across cloud providers. 
Hence, manual emulator development is a tedious process that needs to be repeated for each provider and turns out to be increasingly difficult. The stark reality is that even the most advanced emulator~\cite{localstack} today only covers 95 out of over 240 AWS services, and only with partial API coverage for these services; and no mature multi-cloud emulators exist to our knowledge. 

In this paper, we propose a new approach to building cloud emulators.
We automatically synthesize emulation logic from cloud documentation, and then align the generated emulator with the live cloud to close any remaining gaps.
We observe that publicly available cloud documentation provides a treasure trove of information. No matter how complex a cloud's services may be, their usage and behavior is often described in painstaking detail by the cloud provider. This is fundamental to the cloud's business model, as providers need to sufficiently describe their interfaces to the tenants. 
The incentives are strong for providers to comprehensively document their services and keep the information up to date. 
While poring over these pages is difficult for human developers, Large Language Models (LLMs) have the potential to digest this information and generate emulation code for the described behavior automatically. If successful, this new kind of emulators will not only achieve higher service coverage, but can also easily adapt to service changes, and generalize across providers without multiplying the engineering effort. This version is an extension of our preliminary attempt at the problem~\cite{hotnets-emu}.

We design three steps to \textit{mimic the workflow of a (human) emulator developer}: digesting swaths of information from documentation, coding emulation logic for each resource and its interactions, and testing the emulator against the actual cloud to find and close any gaps. 
While LLMs can help with each step, they can also hallucinate and introduce arbitrary errors. 
Our key observation is that cloud services follow a highly structured style where each resource can be modeled as a \textit{state machine (SM)}.
\textit{transitions} are triggered by API invocations and may further affect the states and transitions of other resources. 
\sys exploits this structure throughout all steps to impose precise constraints and provide high assurance---a neurosymbolic
workflow that ``compiles'' text-based documentation to working code, further tested and refined against the actual cloud as the oracle.  

We call the first step \textit{documentation wrangling}, where the key challenge stems from the sheer \textit{volume} of cloud documentation. 
While LLMs can read/parse voraciously, they can also forget key information after thousands of pages. 
Retrieval-augmented generation~\cite{rag,ICML20-RAG} uses a vector database to help LLMs locate relevant parts of the documentation selectively, but its indexing mechanism is imprecise in nature, using semantic match in the embedding space (e.g., top-k based on the prompt); 
further, understanding the \textit{whole} document is particularly important, as cloud documentation exhibits dense cross-references (e.g., VM definitions refer to NIC attachments, documented far apart). 
\sys develops a \textit{resource index},
that leverages the service and API boundaries, threading through the documentation based on the structure of cloud services for easier LLM consumption.

The previous step indexes and compiles text documentation into a structured  
representation; and the next \textit{code generation} step further compiles any remaining textual descriptions into fully symbolic code. We insist that the LLM generates code following the index structure, translating text to code to fill in ``holes'' in a preprogrammed template that models \textit{a hierarchy of state machines (SM)}. 
Cloud services are complex but modular, allowing the overall task to be decomposed.
This template preprograms key components symbolically. 

Resources are modeled as state machines (SM), and their hierarchical and dependency relationships constrain how resource states can evolve under API operations. 
For example, deleting a parent resource may require first reclaiming its children, while creation and update operations must preserve consistency across dependent resources.
As a result, the LLM only needs to infer, for each API, the corresponding behavior from a small fragment of API description extracted from the large cloud documentation, within this constrained scaffold. 

\looseness=-1
The final step is \textit{program alignment}
which addresses potential gaps that arise from textual ambiguity or under-documentation, testing, and then aligning  
the behavior of the synthesized emulator to that of the actual cloud. It leverages the structure of the template to perform targeted testing, producing high-coverage traces to exercise different program paths. Alignment means that permissible behaviors should produce the same effects in the emulator and the cloud, and forbidden behaviors should fail in both; ideally, failures should also result in identical 
error codes and error messages to assist with DevOps debugging. By testing the effect of these traces in both environments, we can detect divergence, track down the source of errors, e.g., 
to a specific SM implementation, a specific interaction, or even back to the cloud documentation, and perform LLM-assisted program repair. 

These technical designs also contribute key takeaways for building software systems with the assistance of AI. 
(1) LLMs are most effective when used selectively to infer semantics where needed, rather than end-to-end system generation. 
(2) Structured constraints enable generation at scale: symbolic scaffolds that capture architecture and interfaces allow large systems to be synthesized reliably from smaller, composable pieces. 
(3) LLMs can produce plausible but incorrect outputs, and grounding the resulting systems on external oracles and feedback loops can help with continuous validation and alignment.
Our extensive evaluation shows the effectiveness of \sys and validates its high coverage and accuracy over leading, manually-built emulators like LocalStack.

\section{Motivation}
\label{sec:motivation}

Cloud infrastructure is increasingly managed through code.  Whether declarative, IaC/Infrastructure-as-Code configurations (e.g., 
Terraform~\cite{terraform} and similar tools~\cite{cloudformation,pulumi}), or imperative scripts (e.g., Python CLI and SDK programs), these DevOps programs all eventually invoke cloud-level APIs to provision and update resources (e.g., VMs, subnets). 
DevOps programs are safety-critical software, so they must be thoroughly tested and validated. However, testing against the actual cloud is costly, as API invocations are metered and charged; it also introduces high latency (e.g., resource provisioning delay); buggy programs can further endanger the cloud environment. 

To enable high velocity DevOps, emulators that mock up cloud APIs have garnered immense popularity; for instance, they respond to CreateVM() calls by adding a mock VM name, state, and location to internal emulator state, but without actually creating actual resources.  
Emulators are an integral part of the CI/CD pipeline. For instance, a buggy Terraform program may create a VM and its NIC in two different cloud regions; such bugs cannot be detected in the compilation pass~\cite{zodiac}, but API-level emulation will expose this bug before it hits actual infrastructure. 
Obviously, this mock cloud infrastructure has its limitations: it cannot host real workloads or faithfully mimic the cloud's performance characteristics. Nevertheless, DevOps testing
in a frictionless environment is a key value add---the leading AWS emulator, LocalStack~\cite{localstack} (with Moto as the backend), has received more than 60k Github stars.

\subsection{The Pitfalls of Manual Construction}

Existing cloud emulators are manually constructed, which is a sisyphean effort, as the emulation logic often struggles to keep pace with the ever-expanding cloud ecosystem. The cloud is simply too vast. AWS alone has over 240 services, with a single service like EC2 exposing upward of 500 APIs. Furthermore, the cloud is a moving target. AWS announces 3000 changes 
and new features in a year~\cite{awsapichanges}, requiring repeated emulator re-engineering and maintenance to keep abreast. This approach leads to both coverage and accuracy limitations.

\underline{Low coverage:} As Table~\ref{tab:coverage} shows, even popular resources like EC2 have incomplete support, and other resources have even lower coverage. For instance, for Network Firewall---a critical service for cloud security
policies---LocalStack only implements CreateFirewall() but misses 
DeleteFirewall(), UpdateFirewallPolicy(), and
AssociateSubnets() operations. This forces developers to maintain complex and brittle testing setups that use the emulator for some resources and the real cloud for others. Across the {four} services shown, the overall coverage is approximately {35\%}. Further, the penalty of low coverage is combinatorial because an entire DevOps program could fail to execute even if there is one resource/API missing. 

\underline{Low accuracy:} Furthermore, 
subtle behavioral differences between the emulator and the real cloud are common~\cite{icse2025} and can only be revealed by cloud-based testing. For instance, LocalStack allows the DeleteVpc() call to succeed even if the VPC contains an Internet Gateway, while the real AWS  would reject this API with a ``DependencyViolation'' error; calling StartInstances() on a running VM will produce an ``IncorrectInstanceState'' error in AWS, but LocalStack returns a success code instead. 
As the root cause, manual implementation is not infallible, which is compounded by the fact that textual documentation can be ambiguous or incomplete (e.g., using ``etc.'' when describing a set of attributes). 
Left unaddressed, this would  undermine the reliability of the emulator as a testing tool, allowing incorrect code to pass through. 

\begin{table}[t!]
\centering
\caption{API coverage of LocalStack~\cite{localstack}, the leading cloud emulator; even on popular services coverage is under 40\%.}
\vspace{-2mm}
\label{tab:coverage}
\scalebox{0.92}{
\begin{tabular}{@{}lccc@{}}
\toprule
\textbf{Services} & \textbf{APIs} & \textbf{Emulated} & \textbf{Coverage} \\
\midrule
Compute (EC2)    & 756 & 244 & 32\% \\
Firewall (WAF)   &  55 &   30 & 54\% \\
DNS (Route 53)  & 68 &  51  &   75\% \\ 
CDN (CloudFront)    & 167  &  44  &   26\% \\ 
\midrule
\textbf{Overall (subset)} & \textbf{1046} & \textbf{369} &
  \textbf{\textasciitilde35\%} \\
\bottomrule
\end{tabular}
}
\vspace{-4mm}
\end{table}


\underline{Tremendous effort:} 
LocalStack has accumulated over 800 contributors and 10k+ commits across more than a decade of active development. Scaling emulator development to a complex, stateful system that evolves over time is fundamentally hard. 
Covering a new API
requires an engineer to monitor changes, read its documentation, understand its effect on the underlying resource's state, implement that logic
correctly, and write tests; further, APIs do not exist in isolation---the developer must build a mental model of how one resource/API affects other dependent resources and implement their interactions. For instance, 
CreateSubnet() must validate that the referenced VPC exists, record the subnet as a child of that VPC, and later prevent DeleteVpc() from succeeding while the subnet remains. Although these behaviors are documented by the provider, the information is scattered across thousands of pages, so capturing everything completely is a daunting task. 
This labor-intensive process cannot keep pace with a
cloud that adds hundreds of APIs per year. 

\subsection{\sys: The Case for Automated Synthesis}

Fundamental to tackling this task is the ability to pore over many pages efficiently, and translate textual description to running code. Fortunately, LLMs have demonstrated remarkable capabilities in both text comprehension and code generation, and can operate tirelessly to automate this process. 
LLM automation also provides other benefits---it can generalize across cloud providers, and capture service/API changes with periodic relearning, without repeated manual reengineering. 
However, just prompting an LLM to read cloud documentation and generate the emulator code directly performs poorly, exhibiting a wide range of errors---due to context limitations, hallucinations, and the inherent imperfections of documentation. 
To establish intuition behind \sys's techniques, we walk through a simplified example with two resources.

\begin{defn}[Illustrative cloud doc]{defn:spec}
\vspace{-2mm}
\scriptsize{

\noindent\textbf{VPC.} A Virtual Private Cloud resource is a virtual deployment environment, and it can contain several Subnet resources; it exposes the following APIs.

\begin{itemize}[topsep=1pt, itemsep=1pt]
    
    \item \textbf{CreateVpc}(CidrBlock): Creates a VPC with the specified CIDR block. \\ 
    \textit{Inputs:} CidrBlock (str, required); 
    \textit{Returns:} 
    VPC \{VpcId (str), CidrBlock (str)\} or \{ErrCode (enum)\}  

   \item \textbf{DeleteVpc}(VpcId): Deletes the specified VPC. The VPC must not have any
attached \underline{Subnets (\textit{link to the subnet page)}} or other resources.\\
   \textit{Inputs:} VpcId (str, required); 
   \textit{Returns:} Success or \{ErrCode (enum)\}  
   
\end{itemize}
\noindent\textbf{Subnet.} A Subnet has several CIDR blocks  
for a VPC. The CIDR blocks should be valid and within the VPC CIDR range, and they must not conflict  with other Subnets in the same VPC. It exposes the following APIs.
\begin{itemize}[topsep=1pt, itemsep=1pt]
    \item \textbf{CreateSubnet}(VpcId): Creates a Subnet resource within a given VPC. \\
    \textit{Inputs:} VpcId (str, required); 
    \textit{Returns:} \{ErrCode (enum)\} 
    \item $\cdots$
 \end{itemize}
}
\end{defn}

\noindent \textbf{Observation \#1: Unique documentation structure.} We observe that cloud documentation exhibits distinctive characteristics compared to generic textual data. It has a clear structure and clean pagination based on service and API boundaries. As shown in the illustrative example above,  
resource descriptions always start with a resource name, followed by a brief textual description, API inputs, outputs, and error codes. The documentation style is also highly consistent across resources, starting with a comprehensive resource catalog in the beginning pages, followed by marked sections on resources such as VPC and Subnet. This well-defined structure affords a unique opportunity to perform symbolic parsing and retrieval to reduce ambiguity and relieve LLM burdens. 
Instead of RAG-based semantic indexing, each resource is now associated with a feature map, where the key is symbolic (e.g., the resource name) and the value can be textual (e.g., resource description) or symbolic (e.g., API signatures, response formats). Resources also have dependencies---the documentation shows that a Subnet depends on its VPC  because CreateSubnet requires a VpcId. 
These dependencies are symbolic and are represented as edges in this index. 

\begin{figure}[t]
    \centering
\includegraphics[width=0.98\linewidth]{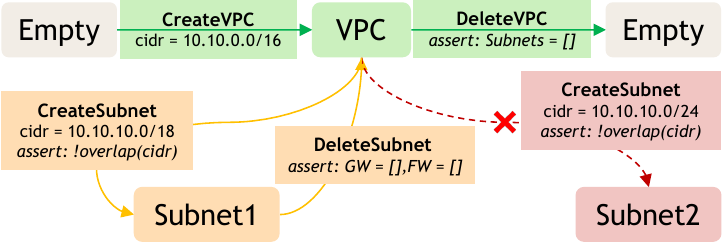}
    \vspace{-1mm}
    \caption{Cloud is a hierarchical state machine allowing inter-SM transitions, resource nodes maintain per-resource state space, with create/destroy API calls to change the state space.}
    \vspace{-3mm}
    \label{fig:sm-hierarchy}
\end{figure}

\looseness=-1
\noindent \textbf{Observation \#2: State machine models.} 
Building upon the above, we further observe that the documentation structure reflects the inherent nature of cloud resources: they are modular and encapsulate state that evolves only through API interactions, which can be modeled in a state machine abstraction. 
For instance, a VPC resource contains CidrBlocks and VpcId as state variables, modifiable via the CreateVpc and DeleteVpc APIs; likewise, a Subnet resource has its own state and state-mutating APIs. The essence of the emulator is to build a hierarchy of state machines, each representing a resource that encapsulates a certain state, further interacting with each other through clearly-defined APIs. This SM hierarchy further follows the dependency edges in the resource index---e.g., a VPC resource is a prerequisite of a Subnet. 
These state machines interact with each other, influencing the state space with APIs as transitions, as shown in Figure \ref{fig:sm-hierarchy}.
As such, we built a generic template that models interacting state machines symbolically.
By grounding code generation in a well-defined structure, we reap the benefits of good abstractions~\cite{liskov}: encapsulating information within modules to separate concerns, and composing modules for programming at scale.

\noindent \textbf{Observation \#3: Cloud-based alignment.} 
Despite the foundational importance of cloud documentation---and providers' incentives to keep it up to date---natural language descriptions remain inherently ambiguous and can be incomplete. Like human readers, LLMs are subject to the same range of ambiguities and potential misinterpretations, as they are trained on natural language corpora. For example, while experienced engineers understand that CIDR blocks within the same VPC must not overlap, others may not readily infer that phrases such as ``conflicts with each other'' in documentation refer specifically to IP address overlap. As a result, emulation logic may fail to enforce such constraints when creating Subnet resources.
To address these limitations, validation against the cloud as a ground-truth oracle is necessary. This requires systematic test case generation that explicitly targets potential ambiguities and edge cases in the documentation. Moreover, the testing framework must detect misalignments between the emulator and the real cloud, such as cases where the emulator reports success while the cloud returns errors. 
\section{\sys Design} 
\label{sec:design}


\begin{figure*}[t]
    \centering
\includegraphics[width=0.95\linewidth]{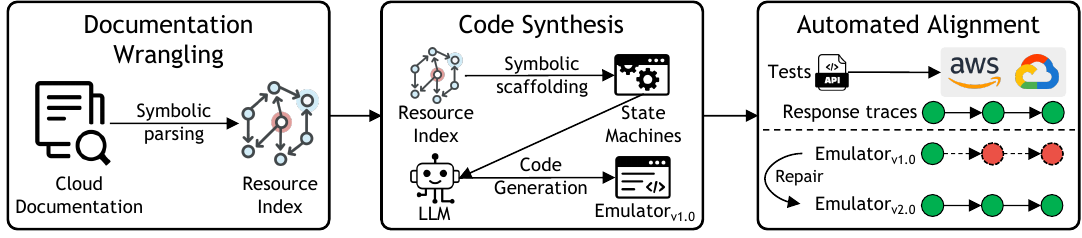}
    \vspace{-2mm}
    \caption{Overview of \sys.}
    \vspace{-3mm}
    \label{fig:system_diagram}
\end{figure*}

Next, we expand on these observations and design three key techniques for documentation wrangling, emulation code synthesis, and automated alignment, as shown in Fig.~\ref{fig:system_diagram}; exploiting the unique structure of cloud resources \& their documents.

\subsection{Documentation Wrangling}
\label{subsec:wrangling} 

\looseness=-1
The cloud is vast, and so is its documentation. The sheer volume presents a challenge to LLMs because of their limited context window---the information needed to emulate a resource is scattered across multiple pages, and dependent resources (e.g., VPC and Subnet) may be documented far apart. Hence, a direct prompting approach does not work well; even retrieval augmented generation (RAG), which converts textual documentation into semantic embeddings for similarity-based retrieval, falls short because of imprecise matches. For instance, a RAG query on \texttt{CreateVPC} parameters may accidentally retrieve content from a semantically-similar API, \texttt{CreateDefaultVPC}, yet these two APIs have different input parameters; 
likewise, a query on VPCs may miss information about dependent resources, such as the state of the Subnet, Security Group, and Load Balancer resources, which are farther apart in the semantic space.
The quality of this retrieval directly impacts emulator correctness, since missing information will lead to incorrect state and state-mutating logic.

\begin{figure}[t]
    \centering
\includegraphics[width=0.95\linewidth]{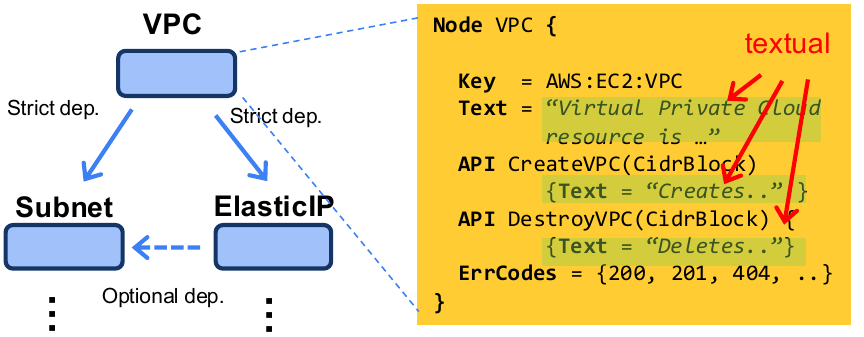}
    \vspace{-2mm}
    \caption{Documentation wrangling builds a cloud resource index. Nodes are resources and edges are (strict or optional) dependencies. 
    }
    \vspace{-3mm}
    \label{fig:index_example}
\end{figure}

How would a human engineer handle this complexity? We observe that the cloud documentation reflects the modular nature of the resources---each is documented in well-defined boundaries and provides cross-references for precise navigation from one resource to its dependent resources (e.g., hyperlinks from VPC to Subnet). 
Further, resource documentation uses a combination of textual description (e.g., the function of a resource), and symbolic definitions (e.g., API signatures and response formats). While the former can be fuzzy and require semantic comprehension, which is imprecise in nature, the latter exists in codified forms and does not require ``guesswork.''  
These unique characteristics allow developers to precisely locate and digest needed information, switching between ``neural understanding'' and ``symbolic parsing'' across different documentation components. 
Hence, our first technique is to construct a \textit{resource index} from raw data in a documentation wrangling phase.  
Figure~\ref{fig:index_example} shows a partial example of the resource index. 
The index captures dependencies among resources, and each node includes both structured fields, such as API parameters, and free-form textual descriptions that are later consumed by the LLM.

\noindent\textbf{Building the index nodes.} 
\sys designs a symbolic parser that decomposes documentation based on service (e.g., EC2), resource (e.g., VPC), and API (e.g., CreateVPC) boundaries. This exploits the structural consistency across documentation pages.
Each API has a fixed documentation structure: a textual description about its functionality, followed by marked subsections on input parameters (names, types, required/optional), response formats in JSON, error codes, and sometimes, usage examples. 
Each resource is an attribute map, with a canonicalized name (e.g., \texttt{AWS:EC2:VPC}) as the primary key. Each resource has a set of attributes that can be indexed hierarchically: e.g., \texttt{AWS:EC2:VPC:CreateVPC} is of the type \texttt{APIDesc}, and its attributes may take symbolic or textual values  (e.g., codified API signature vs. functionality description in text). 

\noindent\textbf{Building the index edges.} 
The second parsing step captures the cross-resource dependencies, e.g., a Subnet lives within a VPC. 
This is done via a traversal of each resource's API definitions to mark whether the APIs presuppose the existence of some other parent resource. 
This also includes optional dependencies, when one resource API references another resource as a non-required parameter, e.g., a Network Interface could be associated with but does not always require an Elastic IP. 

This resource index is better for LLM consumption than raw documentation. 
First, by reorganizing textual documentation in an explicit data structure, we can design algorithms to walk this index and guarantee generation completeness. A full traversal of the index systematically produces emulation logic for all resources and APIs. In contrast, directly consuming raw documentation can easily result in omissions of critical information or logic. Second, the index provides a clean separation between neural and symbolic processing. Information already encoded in structured, symbolic form can be deterministically compiled into code, whereas unstructured textual content is delegated to LLM-based translation. This delineation enables more reliable and interpretable code generation. Third, the modular structure of each node supports incremental generation. Because each node encapsulates a self-contained unit of information, the generation procedure can operate locally without needing to reason about unrelated resources. Finally, the index naturally induces a generation order.  Emulation logic for parent resources are generated prior to their children, ensuring that each child's state and logic can readily reference those from the parent resources.  

\medskip
\subsection{Code Synthesis}
\label{subsec:scaffolding}

Next, \sys synthesizes emulation code to implement the described cloud behavior. Although documentation wrangling significantly improves ease of information digestion, LLM-generated code is never immune from syntactic and semantic errors, e.g., incorrect response fields, missing state variables, and incompatible API interfaces. 
Our key insight is to constrain LLM synthesis using a well-defined, modular abstraction, as implemented in a symbolic framework. This programmatic scaffolding suppresses classes of LLM errors in an otherwise unfettered generation. 
We develop this scaffolding as a one-time effort in a resource-agnostic manner, implementing an abstract model of a cloud resource. The LLM generates code segments from the descriptions for a resource's internals, while the rest is handled symbolically.


\noindent\textbf{The state machine abstraction.} 
Abstractions separate concerns and facilitate composition~\cite{liskov2008powerofabstraction,denning2025abstractions}. 
\sys models each resource as a state machine, where transitions are triggered by API invocations and may further affect the state and transitions of other resources. 
This allows individual modules to be written first and then assembled at scale, which is particularly useful for emulating many interacting resources. For instance, consider the following definition:  
\begin{mdframed}[style=background]
\begin{lstlisting}
/* An abstract state machine */
SM sm { 
    States S; //A collection of state vars
    Transitions T; //Transitions modify state
} 
\end{lstlisting}
\end{mdframed} 
This is simple yet general enough to model a resource without binding the abstraction to resource-specific implementations, which are filled in later when processing each index node. 
For instance, when compiling code for the VPC node, \sys instantiates the abstract definition using VpcId and CidrBlocks as the state; it uses CreateVPC and DeleteVPC as the transitions; it will further add dependent Subnet state machines. 

Specifically, \sys traverses the resource index from the top-level resource (e.g., VPC encloses all other EC2 resources). For each node it visits, \sys compiles its description into a concrete state machine implementation. Hence, compilation proceeds incrementally and modularly for each resource, while temporarily shelving concerns for dependent resources by leaving comments in the code as ``compilation hints.'' For instance, when compiling for the VPC, the resource index will pinpoint the Subnet as a dependent resource; \sys leaves a hint that DeleteVPC must ensure that Subnet resources are empty. After finishing the traversal, 
these incrementally generated ``modules'' are spliced together to form a complete, cloud-wide emulator.

\noindent\textbf{Neurosymbolic compilation.} 
For each node, \sys enumerates attributes in its attribute map and switches between neural and symbolic synthesis based on attribute types.

\textit{State.} We first compile from the index node a set of state variables that are needed to emulate that resource. Our state inference algorithm extracts explicit and implicit state variables using symbolic and neural methods, respectively. 
Explicit state is the union set of variables in a resource that are visible to other resources or DevOps programs, e.g., VpcId and CidrBlocks, thus their corresponding state must be maintained in the emulation. Likewise, Subnet resources depend on their parent VPC, and such dependencies need to be maintained. These externally-visible IO parameters are captured by our compiler symbolically, resulting in a set of state variables for that SM. 
Further, the implicit state is inferred from the textual description and is typically used to maintain runtime state.

\textit{Transitions.} A resource's API descriptions are compiled into a set of state machine transitions. The function signatures and response formats for transition APIs are symbolically defined in the index node. For instance, the \texttt{AWS:EC2:VPC:CreateVPC:IO} is typed as an ``API interface,'' which takes \textit{CidrBlock}, a string type, as input; likewise, its return values follow a fully specified schema: a VPC object containing \texttt{VpcId} and \texttt{CidrBlock}, serialized in XML. 
The transition functionality, on the other hand, is described in prose---the \texttt{AWS:EC2:VPC:CreateVPC:Desc} attribute is a text string extracted from cloud documentation. 


\looseness=-1

\sys applies a ``check-and-fix'' loop to compile this textual description into code. At each iteration, \sys invokes neural models to attempt a translation, and then applies symbolic assertions to check for several invariants. For instance, the  function body must correctly uphold the IO interface, following the expected data formats; 
dependent resources (e.g., Subnet) must be empty before their parent resource (e.g., VPC) is deleted; 
as well as cloud conventions for consistent variable and API naming. 
The loop finishes when the generated function body upholds all invariants.

\subsection{Automated Alignment}
\label{subsec:alignment}


Despite layers of symbolic constraints, an emulator synthesized from documentation alone can still diverge from the real cloud, because natural language descriptions can be ambiguous or incomplete. 
For instance, the CIDR overlap constraint described in \texttt{CreateSubnet} is a case in point: the documentation states that the CIDR block must not conflict with other Subnets in the same VPC, but does not specify the exact validation rule. 
An LLM implementing this constraint may get it partially right, accepting inputs that the real cloud rejects. Likewise, default values from an API response may be missing from documentation, only visible when observing cloud execution. 
This motivates the third stage---treating the real cloud as an oracle, running test cases against both the emulator and the cloud, and using the behavioral differences to drive repair. 
This can be framed as a \textit{automated program repair} problem given a pair of execution traces on the same input program: the cloud's responses are the ground truth, the emulator is the program under repair, and alignment is the process of iteratively finding differences between the observable behavior of the emulator and the behavior of the oracle,
then patching the emulator until its behavior matches the oracle's. 

\textbf{Generating test cases.} 
The complexity in generating test cases lies in the many APIs per resource and the many parameters that can be passed to each API, compounded by the fact that APIs can interleave with each other. For instance, the state space for \(N=100\) APIs with \(p=5\) distinct parameter configurations, and API traces of length \(k=3\), is on the order of \((N \cdot p)^{k}=125\) million. 
Although state space explosion is a classic problem in program analysis, the cloud further poses additional scalability challenges because API latencies are much higher and more costly than local programs. 
Hence, the test case selection needs to be wary of limited time and monetary budgets; an exhaustive enumeration does not scale.

We use a neurosymbolic approach to test case generation that combines symbolic execution and LLM-based synthesis. Our primary target is the API transition body, which is generated by the LLM. We further isolate each API and perform symbolic execution on its implementation as the basic unit, resulting in a set of path constraints that would exercise each branch in the function.
We use these path constraints as coverage targets for test generation alongside the total state touched (based on resource attributes read or assigned).
We then invoke an LLM to pick which APIs to test and in which order, while asking the LLM to prioritize realistic invocation orders. Specifically, we ask the LLM to generate CLI or Terraform programs that would trigger the underlying APIs, instead of generating raw, API-level traces, because these DevOps programs reflect real-world API usage.

\looseness=-1
\textbf{Localizing root-cause from noisy traces.}
For each test case, we execute the program in both the emulator and the cloud, record and analyze the traces. 
The goal is to automatically detect where the emulator diverges from the cloud. The challenge arises from noisy traces, especially from the cloud. 
Many fields, such as server-generated IDs and timestamps, differ across runs and lead to discrepancies without a semantic difference. 
In addition, cloud APIs have latencies that vary depending on the API, and this results in reordered operations in the cloud side. 
\sys aligns the two traces by matching the same API call on the same logical resource, such as matching both runs of \texttt{CreateVpc} for the same VPC name. 
We align the two noisy traces to identify corresponding operations, then localize the earliest root-cause discrepancy while suppressing downstream noise, and finally map the misbehaving API to the exact generated code region for repair.

\textbf{Patching.}
After localization, the remaining task is to repair the emulator so that its behavior matches the real cloud. 
At a high level, we classify the surviving API response discrepancies into four categories: \texttt{MISSING\_FIELD}, \texttt{EXTRA\_FIELD}, \texttt{VALUE\_MISMATCH}, and \texttt{STATUS\_MISMATCH}. 

\looseness=-1
We first perform symbolic passes that attempt to resolve discrepancies whose fixes are directly observable from the reference trace. 
The most important of these is \emph{response seeding}: when a field’s correct value is visible in the reference trace.
For example, if the reference trace shows that \texttt{CreateSubnet} always returns \textit{mapPublicIpOnLaunch: false} as a default but the emulator omits the field entirely, response seeding injects the assignment symbolically.
A second symbolic pass repairs formatting mismatches, such as when the emulator returns a scalar where the reference returns a list. 
For example, a \texttt{CreateSubnet} response returning \textit{cidrBlockAssociationSet} as a plain string rather than a single-element list is a common instance of this kind. 
Together, these two passes resolve a fraction of discrepancies mechanically and run in seconds, making them worth applying before any model invocation. 

For discrepancies that symbolic patching cannot resolve, 
we extract the full method source alongside the structured discrepancy list and use LLM-based repair. This returns a replacement written at the exact line range identified by the AST index.
The patcher re-runs the representative case after each attempt and re-localizes surviving findings. 
If the discrepancy set shrinks, the retry budget extends, rewarding progress; if the patcher makes no progress across retries, 
we explore patches that cross APIs, such as a \texttt{DescribeSubnets} response that must join state from both the subnet store and the VPC store to populate nested association fields, or a \texttt{DeleteVpc} that requires coordinated cleanup across multiple dependent resource stores. 
All patches are applied to a separate copy of the generated code, and the patched emulator is re-evaluated in subsequent alignment runs to close the loop. 
\vspace{-1mm}
\section{Evaluation}
\label{sec:eval}

Next, we evaluate \sys to answer the following research questions: 
\vspace{-0.5mm}
\begin{packeditemize}
    \item[] \textbf{RQ1}: How effective is \sys in emulating the cloud, compared to the leading emulator, LocalStack?  
    \item[] \textbf{RQ2}: How effective are the three key designs in improving \sys performance? 
    \item[] \textbf{RQ3}: What is the overhead of synthesis and alignment? 
    \item[] \textbf{RQ4}: How well does \sys generalize across clouds? 
\end{packeditemize} 
\vspace{-1.5mm}

\subsection{Methodology} 

\noindent \textbf{Implementation.} 
We implemented \sys in 12.4k lines of code in Python: 1.2k for documentation wrangling, 7k for scaffolding and code synthesis, 
and 4k for automated alignment. 
Our main evaluation target is AWS's EC2, one of the most complex and widely-used services; to demonstrate generalizability, we report results on emulating GCP's Compute Engine. The synthesized emulator (88k and 65k lines of code for EC2 and GCE, resp.) runs as a local instance, behind a Flask server gateway. DevOps programs, whether imperative CLI scripts or declarative Terraform configurations, can issue API calls into this gateway for emulated execution.

\noindent\textbf{Metrics.} The key metrics to measure an emulator are coverage and fidelity. Coverage is measured by the number of emulated APIs divided by the total number of APIs in a cloud. Fidelity measures the similarity between an emulator's response to a test case (one or more API invocations) and the response from the actual cloud. We do not require identical responses, as simulated timestamps, resource IDs, and other fields, do not affect functionality and can be different across actual cloud executions as well. We measure fidelity using three levels. 
\vspace{-2mm} 
\begin{packeditemize}
    
\item \textit{L0 (Error Class):} the emulator and the cloud respond with identical error codes for all APIs involved in a test case, but we do not examine other return values. 
\item \textit{L1 (Response Fidelity):} Passes L0 (ErrorClass) and also, key response fields (e.g., CidrBlocks) are identical. 

\item \textit{L2 (State Consistency):} Passes L1 (IO behaviors) and also, the emulated state is equivalent to that of the cloud. 
\end{packeditemize}
We measure fidelity using a suite of test cases (e.g., CLI commands or Terraform configurations). 
Each test case produces an API trace in the emulator and the cloud, and we consider the emulator to pass a fidelity level (L0-L2) if and only if all API responses and intermediate states pass that level.

\subsection{Emulation Effectiveness: Versus AWS} 
\label{sec:eval:alignment}

\begin{figure}[t]
  \centering
  \includegraphics[width=\columnwidth]{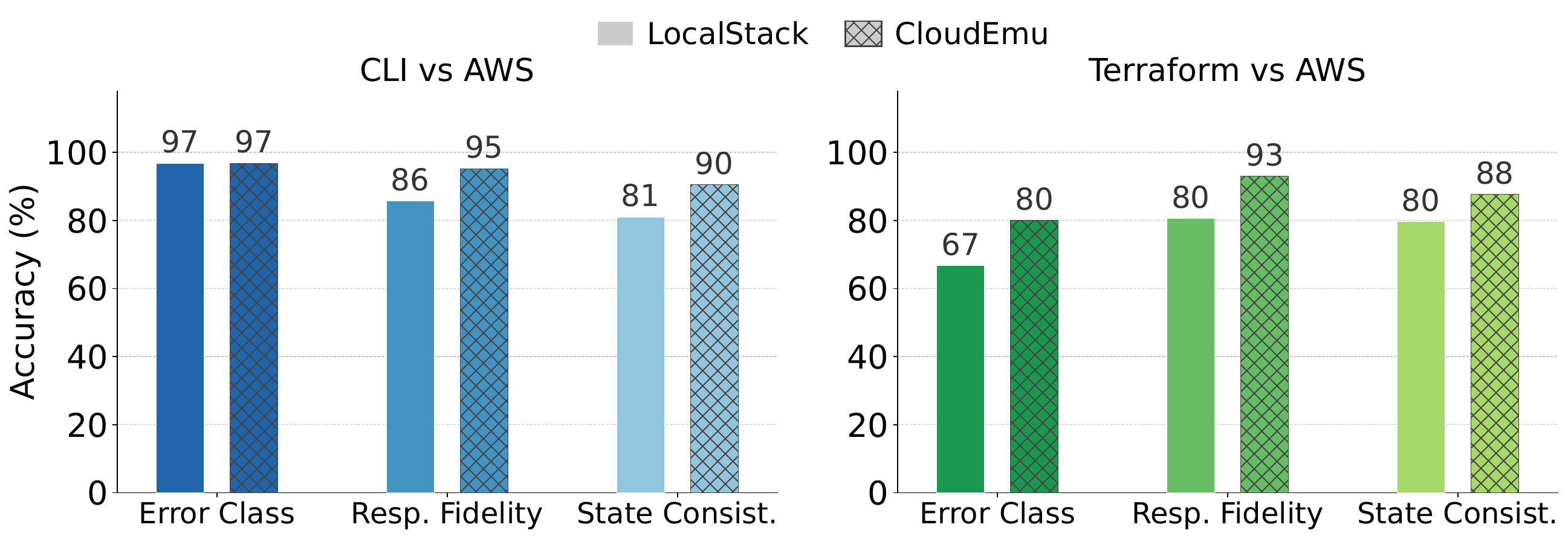}
  \caption{LocalStack and \sys scored against real AWS traces (50 CLI + 50 Terraform cases).}
  \label{fig:ablation:aws}
\end{figure}

\begin{figure*}[t]
  \centering
  \begin{minipage}[t]{0.48\textwidth}
    \centering
    \includegraphics[width=\textwidth]{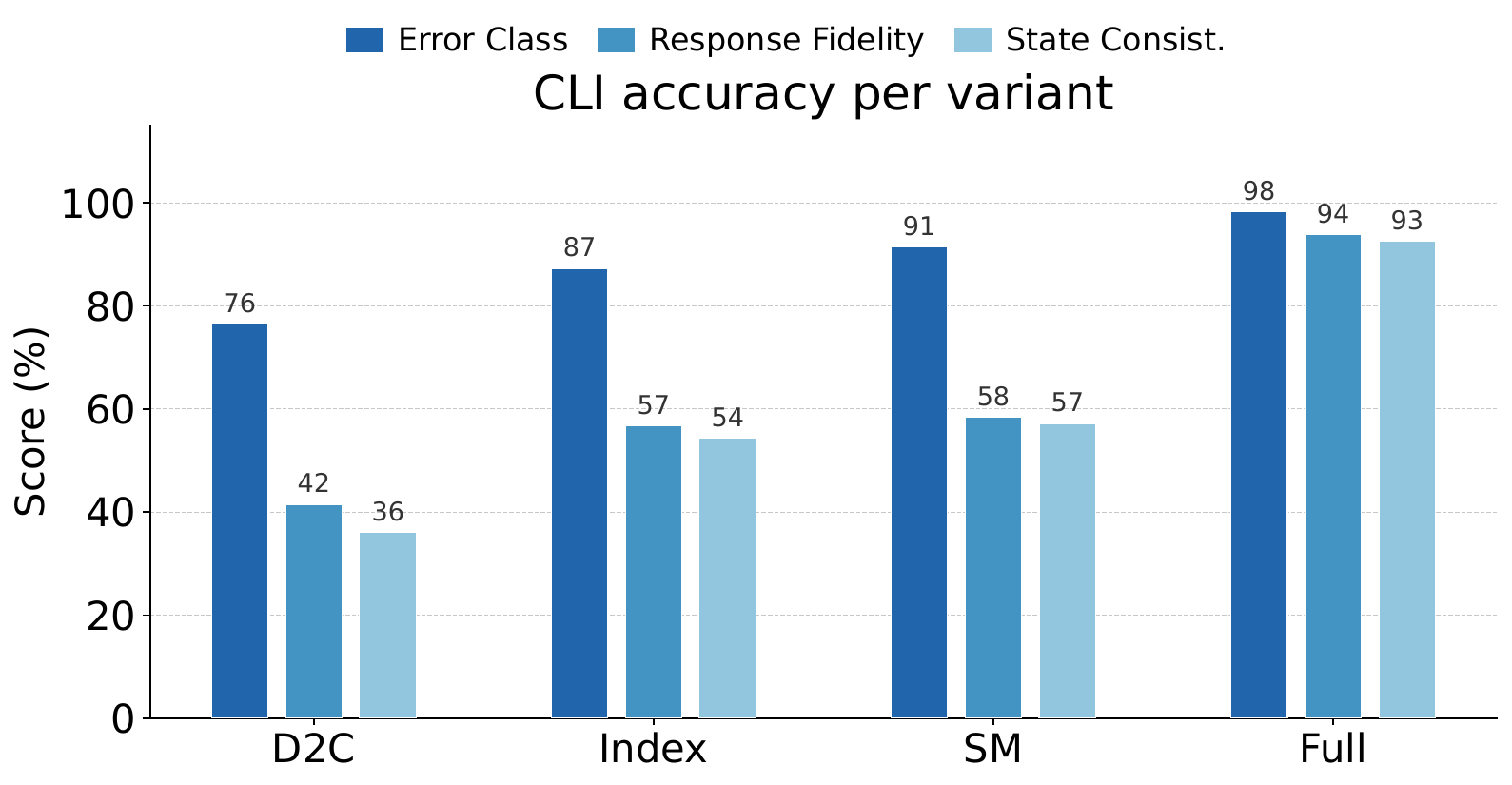}
    \vspace{-6mm}
    \caption{
      CLI ablation results (267 cases).
    }
    \vspace{-2mm}
    \label{fig:ablation:results:cli}
  \end{minipage}
  \hfill
  \begin{minipage}[t]{0.48\textwidth}
    \centering
    \includegraphics[width=\textwidth]{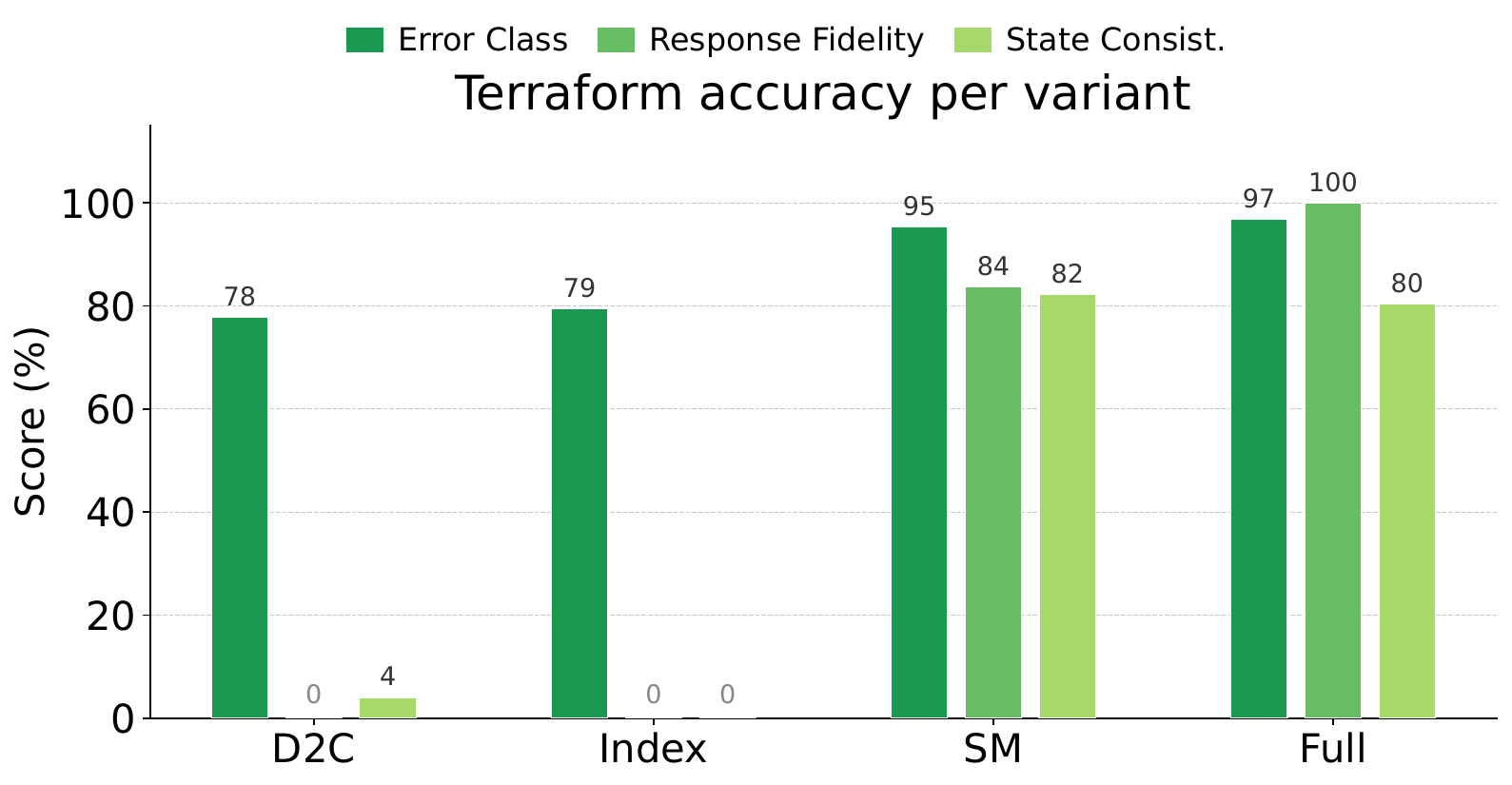}
    \vspace{-6mm}
    \caption{
      Terraform ablation results 
      (63 cases).
    }
    \vspace{-2mm}
    \label{fig:ablation:results:tf}
  \end{minipage}
\end{figure*}

We start by measuring emulation effectiveness of \sys against the leading baseline, LocalStack, using AWS as the ground truth. For the entire EC2, \sys achieves 100\% API coverage across these randomly chosen cases, whereas LocalStack achieves only around ~70\% case coverage. We further choose the remaining cases (with an additional 20 randomly chosen cases) to align and generate \sys (such that the test set is not used for alignment). For fidelity, 
as we can see from {Figure~\ref{fig:ablation:aws}}, \sys consistently outperforms LocalStack at all fidelity levels. 
Generally, Terraform cases are harder to emulate accurately due to the general depth of the number of APIs per case, with state dependencies. 
Despite this, \sys achieves higher scores than LocalStack for both Response fidelity and State consistency. 

As expected, Error Class parity is the easiest dimension: both systems match AWS error behavior on CLI cases, but the gap widens on Terraform, where LocalStack accepts invalid configurations and parameters and allows Terraform programs with dependency violations (e.g., deleting a Subnet that still has resources attached). Similarly, for the other two task categories, a primary reason Localstack fails is that it truncates responses with stripped down fields compared to AWS. As an example, on the \texttt{RevokeSecurityGroupEgress} API, LocalStack returns only \textit{{"return": "true"}}, whereas AWS (and \sys) return the full \textit{revokedSecurityGroupRuleSet} containing the revoked protocol, port range, and CIDR. \sys is able to leverage alignment to generalize and remove certain classes of errors, which in turn gives gains over a fresh set of test cases as well. Table \ref{tab:ablation:patch_examples} and \ref{tab:ablation:patch_dist} cover examples of the kinds of errors and their proportion under
\sys. A majority of the errors were due to missing fields in the response (i.e., the backend logic on adding blocks based on input parameters, like IPv6 state, is flawed in certain places, or at times, nested fields aren't serialized appropriately).

\subsection{Ablation Study: Versus LocalStack}
\label{sec:eval:ablation}


\begin{table*}[]
\centering
\small
\caption{Examples of behavioral corrections introduced by LocalStack-guided alignment in \sys-LS and by AWS-guided alignment in \sys.}
\label{tab:ablation:patch_examples}
\renewcommand{\arraystretch}{1.1}
\begin{tabularx}{\linewidth}{@{} l X X @{}}
\toprule
\textbf{Mismatch class} & \textbf{Example in \sys-LS Alignment} & \textbf{Example in \sys Alignment} \\
\midrule
\texttt{MISSING\_FIELD}
& \footnotesize\textit{CreateLaunchTemplate}: added missing nested response fields that were expected by the LocalStack execution trace.
& \footnotesize\textit{CreateVpc}: added missing fields such as \textit{CidrBlockAssociationSet}, IPv6 state, and owner ID to match AWS responses. \\

\texttt{STATUS\_MISMATCH}
& \footnotesize\textit{CreateDhcpOptions}: corrected whether the call should return success or an error.
& \footnotesize\textit{DescribeSecurityGroups}: corrected the call to return success, matching AWS behavior for a newly created security group. \\

\texttt{VALUE\_MISMATCH}
& \footnotesize\textit{CreateSecurityGroup}: corrected response values echoed back to the caller to match LocalStack behavior.
& \footnotesize\textit{DescribeKeyPairs}: corrected the returned \textit{KeyFingerprint} to match the fingerprint computed by AWS. \\

\texttt{EXTRA\_FIELD}
& \footnotesize\textit{CreateSubnet}: removed or reshaped extra response fields so that the returned object matched LocalStack more closely.
& \footnotesize\textit{DescribeInstances}: removed internal state fields that appeared in the emulator output but are not present in AWS responses. \\
\bottomrule
\end{tabularx}
\end{table*}

\begin{table}[t]
\centering
\small
\caption{Distribution of patch attempts by mismatch type in the LocalStack- and AWS-guided alignment stages.}
\label{tab:ablation:patch_dist}
\renewcommand{\arraystretch}{1.1}
\begin{tabular}{lcc}
\toprule
\textbf{Mismatch type} & \textbf{\sys-LS} & \textbf{\sys} \\
\midrule
\texttt{MISSING\_FIELD}   & 16.3\% & 35.0\% \\
\texttt{STATUS\_MISMATCH} & 13.5\% & 30.0\% \\
\texttt{VALUE\_MISMATCH}  & 32.0\% & 22.5\% \\
\texttt{EXTRA\_FIELD}     & 38.2\% & 12.5\% \\
\bottomrule
\end{tabular}
\end{table}

Next, we evaluate \sys using a larger set of test cases (330 in total, including 267 CLI scripts and 63 Terraform configurations). To save cloud cost (and time to cleanup), we perform this ablation study using LocalStack instead of the actual AWS as the alignment target to evaluate the performance improvements from each technique. 
We consider several variants of \sys: 
\vspace{-2mm}
\begin{packeditemize}
    \item \textit{Direct to code (D2C)}: Directly prompting LLMs to read cloud documentation and write the emulator.  
    \item \textit{Index}: Uses our cloud resource index, while leaving resource-level code generation to LLMs.
    \item \textit{SM}: Further adds symbolic generation based on the state machine abstraction, but without alignment. 
    \item \textit{Full}: All techniques, further aligns the emulator against the oracle, which in this subsection is LocalStack. And we denote this system variant ``\sys-LS''.
\end{packeditemize}

\noindent\textbf{Coverage:} 
Na\"{i}ve baseline (D2C) reaches a coverage of 66\%, failing to cover all APIs. This is because comprehensive coverage fundamentally requires a precise resource index:
Even though LLMs can read efficiently, token-based text parsing is not the best way to capture symbolic API definitions, and could lead to missed APIs. The Index, SM and Full variants on the hand achieve perfect coverage.  

\noindent\textbf{Accuracy:} 
We then measure the improvement in emulation accuracy at each step, and show the results in Figure~\ref{fig:ablation:results:cli} and \ref{fig:ablation:results:tf}. 
Index and SM represent the biggest improvements across both modality of cases, because they address the two key problems with LLM generation: long context and arbitrary errors. Adding the resource index improves CLI Error Class accuracy by 11\% and response fidelity by around 15.3\%, because it gives the LLM fine-grained, per-resource context rather than asking it to ingest the full documentation at once; the generated handlers are longer, more complete, and correctly populate more response fields.

With SM, we found that the biggest accuracy boost comes for Terraform cases with a 15\% gain in Error Class accuracy and a big 80\% gain in the Response Fidelity. The state-machine scaffold enforces correct API signatures and response schemas before any code is generated, thus avoiding LLM-generated TF handlers returning malformed XML. Moreover, a lot of the dependency violations existing in Localstack programs get prevented since SM naturally follows a dependency graph. For CLI calls, the change is negligible since a lot of the CLI operations aren't passing any checkable parameters; for the ones passing these parameters most of the problems are because of field initializations and internal logic, not parsing or response creation. 

Initially, 29.9\% APIs for VPC had signature mismatches, and 41.8\% APIs had response format errors; after the SM is applied, both reduce significantly, with signature mismatches dropping to 9\% and response formats getting completely fixed (with an exception of extra fields for some cases). 
The alignment phase resulted in a small improvement for error class accuracy and Terraform cases in general, meaning that the documentation is already quite useful to bootstrap the emulator. For the CLI cases, response and state fidelity improved significantly through alignment, because a majority of those errors were due to incomplete state initialization (e.g., default values such as list of existing images at VM startup).
Such information is barely covered in the documentation, thereby needs to be properly aligned via an oracle. 
This shows that alignment is a vital step, as mismatches can occur in purely documentation-based generation. 

We show two examples where alignment is helpful: i) missed/incomplete validation checks in CreateSubnet, which for reference do not check whether the Subnet being created lies in a different Availability Zone as its parent VPC, since the documentation doesn't state this directly; and ii) incomplete state initialization for fields in the resource state, like for \texttt{RunInstance} with a ami\-id passed in, real cloud queries through the predefined templates to validate which the emulator has incomplete knowledge about. Similarly a bunch of optional fields like (\textit{CarrierIp},
\textit{NetworkBorderGroup}, etc) aren't initialized correctly for \texttt{AllocateAddress}. Examples of more such errors and their proportion are covered in Table \ref{tab:ablation:patch_examples} and \ref{tab:ablation:patch_dist}, respectively, under \sys-LS, with the major proportion being extra fields in responses.

\subsection{Overhead Analysis}  
\label{sec:eval:cost}

\begin{table}[]
\centering
\small
\caption{Normalized cost of \sys code synthesis and automated alignment on AWS EC2.}
\label{tab:cost_summary}
\renewcommand{\arraystretch}{1.1}
\begin{tabular}{lcc}
\toprule
\textbf{Stage} & \textbf{Time (s) } & \textbf{Tokens (k)} \\
\midrule
Synthesis          & 52 / resource & 478 / resource \\
\sys-LS Alignment  &  21 / case   & 3.4 / case \\
\sys Alignment     & 126 / case   & 4.1 / case \\
\bottomrule
\end{tabular}
\end{table}

Next, we measure the overhead of \sys. The resource index takes around 10s to build for EC2, and incurs the least amount of overhead across the three steps. This is because building the index is a deterministic step using symbolic parsers, and although the documentation is long, the processing is efficient, and the parser can extract the pages quickly.
As for the SM method, the symbolic framework is written in around 7k lines of code, only a fraction of the entire emulator code. 
Everything else is generated by the LLM while being in the constraints of this framework. This is a massive amount of workload reduction compared to manually reading and implementing all cloud emulation logic by hand. The entire emulators for AWS and GCP consisted of 88k and 65k lines of code, spanning 90 and 91 resource types, respectively. 

Per resource, approximately 16
LLM calls are needed to generate the code; smaller resources with fewer APIs
took around 10 calls, and larger resources (e.g., \texttt{instance.py}: 3,724
lines) took around 22 calls. We have tried this task using three different LLMs: GPT 5.1, GPT 5.2-mini, and GPT 5.2-codex. 
We found that weaker or newer models do not result in a visible accuracy difference, because our symbolic method can boost the effectiveness of this generation, so that even smaller models suffice for this task. 
Though older, smaller models like GPT 4o suffer from limited expressibility and don't generate entire API logic at once correctly.
Overall, as shown in Table~\ref{tab:cost_summary}, the synthesis takes roughly 1.3 hours.

Next, we measure the alignment overhead. This is dependent on the alignment target. In Table~\ref{tab:cost_summary}, we show the result on aligning against LocalStack versus the real AWS cloud. 
As expected, per program the amount of time it takes to align against the oracle LocalStack is $6\times$ faster (21~s/case vs.\ 126~s/case) because it runs locally on device. For reference, in the 100 cases run in \ref{sec:eval:alignment}, we found 24 APIs with
discrepancies across 48 alignment cases, yielding 41 patch operations. In contrast, for AWS, each alignment run takes longer, as \sys needs to rerun against the cloud, obtain traces, and cleanup for the next run.
\begin{figure}[t]
  \centering
  \includegraphics[width=\columnwidth]{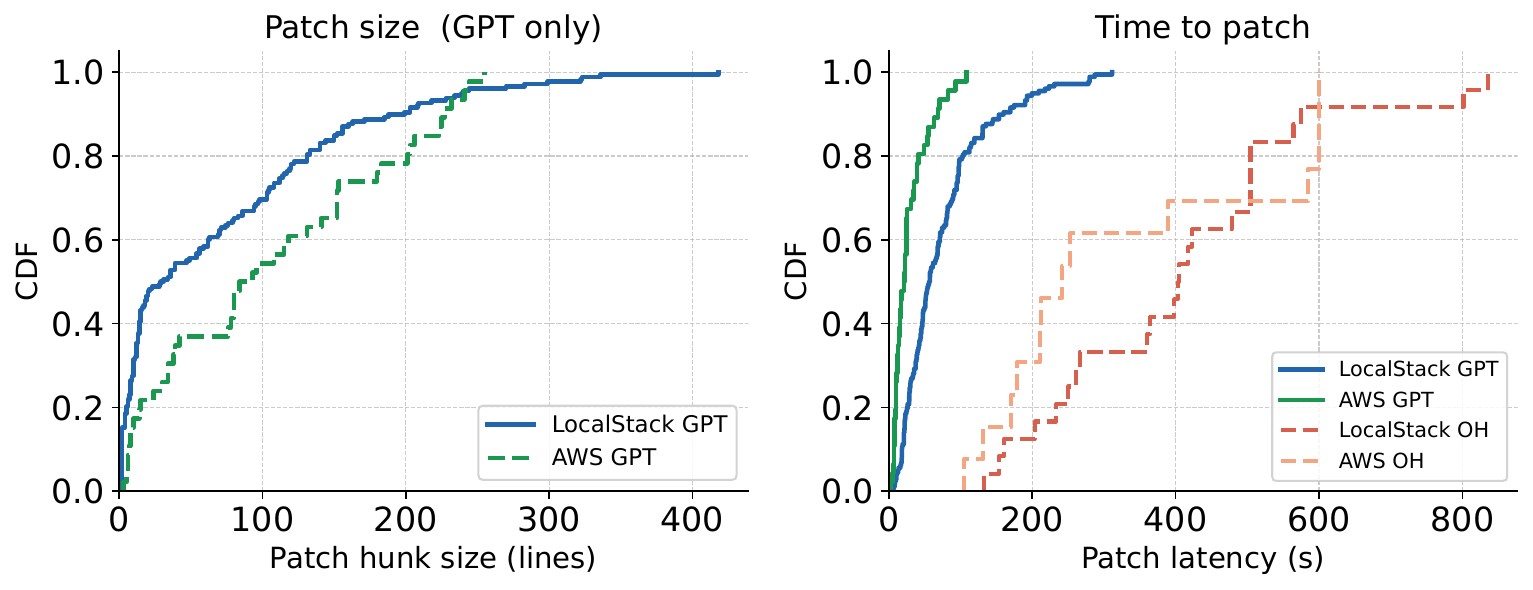}
  \caption{CDFs of patch size (left) and latency (right) for GPT patching attempts; verified OpenHands escalations are included in the latency panel only.
  }
    \vspace{-2em}
  \label{fig:ablation:aws_patch_cdf}
\end{figure}


\looseness=-1
Additionally, alignment cost depends not only on the target oracle but also on the discrepancy types uncovered, shown in Table~\ref{tab:ablation:patch_dist}.
Different classes of errors demand different repair effort, which directly affects patching overhead. We therefore next break down overhead by error type.
Against LocalStack, \texttt{EXTRA\_FIELD} and \texttt{VALUE\_MISMATCH} dominate, reflecting structural over-generation that LocalStack can detect because it returns those fields itself, 
whereas against AWS, \texttt{MISSING\_FIELD} and \texttt{STATUS\_MISMATCH} become dominant, exposing semantic gaps that LocalStack cannot surface because the latter shares the same omissions. 
Figure~\ref{fig:ablation:aws_patch_cdf} shows the CDFs of patching effort for all GPT and coding agent (OpenHands/OH~\cite{wang2025openhandsopenplatformai}) attempts. Half of all patches touch fewer than 84 lines and complete in under 30 seconds, but the distributions have long tails: the top 10\% exceed 246 lines and 180 seconds, reflecting cases where correcting existing values or removing fields requires touching interconnected logic. These stalling cases are further escalated to a full coding agent (OH).

\begin{figure}[t]
  \centering
  \includegraphics[width=\columnwidth]{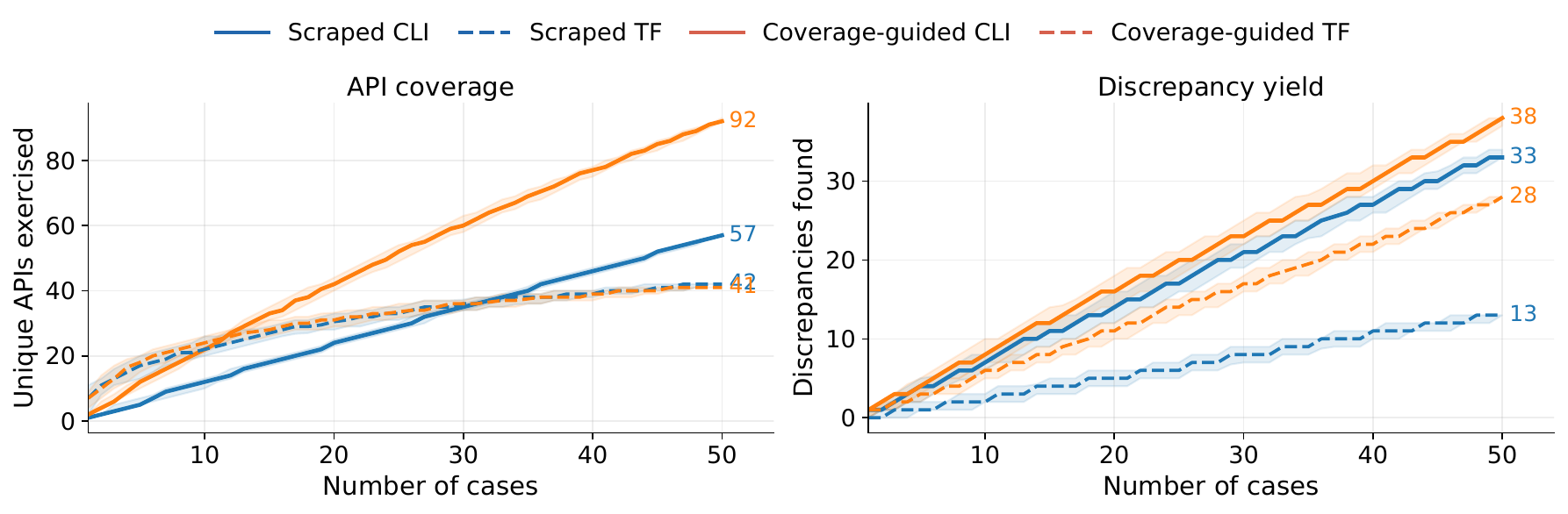}
  \caption{Cumulative unique APIs exercised and discrepancies
    found for scraped vs.\ coverage-guided cases}
  \vspace{-4mm}
  \label{fig:alignment_efficiency}
\end{figure}

Beyond the per-discrepancy patching cost, overall alignment efficiency also depends on the strategy used to discover discrepancies. 
Targeted exploration can uncover alignment gaps with fewer test cases and thus lower total cost. 
We measure the effectiveness of coverage-guided alignment, targeting emulator branches left uncovered by the benchmark suite, to find alignment gaps more efficiently.
We generate 50 CLI and Terraform \textit{coverage-guided cases} each via a greedy set-cover over static branch targets extracted from the pre-aligned emulator source code.
These are synthesized by GPT 5.2-codex, picking branches with low confidence scores (assigned by LLM during code generation) to prioritize cases where the generation was unclear.
Comparing this with the strategy of scraping random cases, Figure~\ref{fig:alignment_efficiency} shows cumulative unique APIs exercised and discrepancies found as the number of cases executed increases (averaged over 200 random orderings for the scraped cases). 

The coverage-guided strategy helps us uncover much more discrepancies per case than random scraped cases.
At a budget of 50 cases each, coverage-guided cases exercise 79 unique EC2 API methods compared to scraped CLI cases(88\% more), including 37 APIs absent from the scraped suite entirely.
The discrepancy yield is also higher: for Terraform, coverage-guided cases surface 66\% discrepant cases compared to 27\% for scraped, a $2.5\times$ improvement. The two strategies expose complementary fault classes: scraped cases find more field-value mismatches ($2.2\times$) on well-exercised APIs, while coverage-guided cases expose $2.1\times$ more Error Class mismatches on parameter paths the scraped suite never reaches, surfacing 32 distinct AWS error codes versus 19 for scraped, with 22 unique to the coverage-guided suite.

\subsection{Translating \sys to GCP}  
\label{sec:eval:gcp}

\sys\ transfers to GCP Compute with minimal provider-specific adaptation.
The synthesis, patching, and alignment loop remains unchanged; only the template layer must encode provider-specific interface conventions.

\noindent\textbf{Adapting the pipeline to GCP.}
The differences between AWS EC2 and GCP Compute are architectural
rather than algorithmic. GCP resources are identified by string names
rather than prefixed IDs, all mutating calls return \texttt{Operation}
objects rather than direct resource responses, labels replace tags,
and error responses use JSON rather than XML\@. We encoded these
differences as GCP-specific templates in the symbolic generator. 
The OpenHands patching and alignment stages remain unchanged.

\begin{table}[t]
\centering
\small
\caption{Accuracy on Terraform cases over GCP Compute resource cases}
\label{tab:gcp_tf_results}
\renewcommand{\arraystretch}{1.1}
\resizebox{\columnwidth}{!}{%
\begin{tabular}{lccc}
\toprule
\textbf{\sys-GCP}  & \textbf{Error-class} & \textbf{Resp. Fidelity} & \textbf{State Consist.} \\
\midrule
before alignment & 0.644 & 0.571 & 0.545 \\
after alignment & 0.956 & 0.857 & 0.734 \\
\bottomrule
\vspace{-2em}
\end{tabular}
}
\end{table}

\noindent\textbf{Results.}
We evaluate 45 Terraform cases (spanning different numbers of resource blocks and dependencies), across 18 GCP Compute resource types, scraped from official registries and repos. 
Similar to AWS, we evaluate these cases on three metrics against real GCP.
As shown in Table~\ref{tab:gcp_tf_results}, the \sys-GCP after alignment achieves high scores across all metrics.
An example of error-class miss is a disk resource gated on a Windows OS license not active on our billing account; GCP returns \textit{403
Permission\_Denied} while the emulator succeeds; this is orthogonal to resource wiring.

Two general classes accounted for the failures at all levels.
\textit{i) Incorrect resource identifiers}: regional backends stored the region as a full resource URL (\textit{https://.../regions/us-central1})
but GET lookups on the short name, and \textit{selfLink} fields used the wrong collection name (\textit{RegionBackendServices} instead of \textit{backendServices}), causing Terraform's read to fail.
\textit{ii) Incomplete target validation}: forwarding rule insertion rejects some valid VPN configurations as \textit{service\_attachment} and \textit{vpn\_gateway} were not included as valid target types.
The one remaining failure, \textit{google\_compute\_network\_peering}, requires bidirectional state updates not modeled by the per-resource scaffold. These error types get fixed by alignment, and the emulator gains roughly 10-20\% on the different metrics.

\noindent\textbf{Takeaway.} 
The pipeline generalizes to GCP with only template-layer changes and configurations; synthesis, patching, and alignment stages transfer unchanged.
The high scores across the three metrics confirm that the core insight of the paper holds across providers:
typed scaffolding and centralized state are sufficient to capture cross-resource wiring, and the alignment loop reliably closes the gap between generated structure and provider-specific behavior. 
\section{Discussion} 



We discuss the limitations in the design of \sys as well as potential future use cases for this direction.  


\noindent \textbf{Limitations in performance and failure emulation.} Currently, emulators only respond to API invocations but do not simulate the performance characteristics like invocation latency, rate limits, \& network latencies. There are applications that may require this level of simulation, e.g., for performance testing or chaos engineering. For these use cases, LocalStack and \sys need to be extended to consider performance or failure profiles in addition to API fidelity. Likewise, simulation of semantic properties, such as concurrency tests or consistency properties, are not handled by existing emulators. 

\noindent \textbf{Limitations of third-party implementation.} Thus far, both LocalStack and \sys take a third-party approach to implementing cloud emulators. This is because cloud providers themselves prioritize building actual cloud features to stay competitive in the market, rather than emulating them. 
Hence, emulators are primarily developed by third-party developers based on the cloud's documentation. However, if the cloud providers were to implement emulators themselves, they could take two approaches. One approach is to reuse their existing codebase for the actual cloud service, and then derive a lightweight emulator from the implementation. However, actual implementations are coupled with hardware details and are distributed systems, which may necessitate a high amount of reengineering work to create a local mock version. The other approach they could take is to rely on documentation, whether public API documentation or private design documentation. \sys could help with this latter approach with automated synthesis.

\noindent \textbf{Cloud gym:}
This emulation framework can also act as a playground for learning and testing cloud services for AI agents. There has been a recent line of work on building AI agents for cloud management~\cite{cloud-agent, aiopslab, jha2025itbenchevaluatingaiagents, yang2025nsync, yang2026ambigiacmultileveldisambiguationinteractive},
with the goal of eventually automating DevOps engineering. 
To train such agents, we need a high-fidelity cloud gym, for example, for reinforcement learning, that provides a no-cost, zero-risk environment for generating feedback and learning data. Such a gym could also be used to validate agent behavior before deployment on the real cloud.

\noindent \textbf{Documentation engineering:} By analyzing the specifications, we can detect potential design flaws and anti-patterns. For instance, a modify() call that requires a long and complex chain of actions updating multiple dependencies across resources may indicate a poorly designed API; or, documentation that consistently leads the AI to generate incorrect logic may be flagged as ambiguous and in need of refinement. This will improve API and documentation engineering~\cite
{google-api-engineering}. The synthesized emulator can also be used as an executable specification than textual documentation. 

\noindent \textbf{Multi-cloud emulation:} Our approach is provider-agnostic and can generalize to any cloud backend. By consuming different cloud providers' documentation, we can generate a standardized formal model for all of them, and generate a ``universal emulator'' for testing multi-cloud DevOps programs. Our approach also enables formal, automated comparisons of equivalent services---e.g., whether Azure's CreateVM() requires the same dependency checks as AWS's RunInstance() in AWS---and can help improve cross-cloud portability.


\noindent \textbf{Quantifying cloud complexity:} The resource index comprises a graph of interacting state machines. This provides objective metrics (e.g., number of nodes, edge density) for a quantitative analysis of cloud service complexity. This allows for comparisons, for example, between the complexity of AWS Lambda and Azure Functions, and could assist cloud providers to modularize their resources. 

\noindent\textbf{Direct generation using LLMs} As the frontier LLMs evolve, there have been increasing attempts to build alternatives to the existing emulators (\cite{localstack}) by one-shot synthesis using coding harnesses. These implementations (e.g., Floci~\cite{floci}) suffer from severe fidelity issues, rendering them impractical for emulation and requiring lot of manual effort to improve fidelity.

\section{Related Work} 
\label{sec:related}




\noindent\textbf{Specification mining.} Our work builds on a rich history of techniques for automatically extracting API specifications 
~\cite{doc2spec,mining_preconditions,generating_oracles, oracle_2,learning_logic_reps}, especially those that leveraged LLMs to directly translate informal natural language comments and documents into checkable assertions~\cite{llm2spec_fse24, c2s_fse20}, temporal properties~\cite{llm_temporal_properties, nl2spec}, and other formal specifications~\cite{llm_synthesis_of_specs,llm_synthesis_of_specs_2, specgen_ase24}. 
While we draw heavily from these advances, most of them stop at inferring invariants for existing implementations, rather than generating end-to-end emulation code.

Within the domain of spec mining, our work is most related to active automata learning, or ``model learning''~\cite{model_learning, model_learning_2}, where an algorithm interactively queries a black-box system to infer a state machine model of its behavior using traces or documentation. The most related work, Hermes~\cite{hermes_fsm} synthesizes FSMs from network protocol documents to enable security analysis, while other works have focused on automated attack synthesis~\cite{attack_syn_fsm}, automated testing~\cite{auto_test_fsm}, or detecting bugs in protocol implementations~\cite{model_bug_2, model_bug}. 
To the best of our knowledge, our system is the first to use LLMs to translate API documentation into a complete, executable state machine model, specifically to emulate complex cloud services.



\noindent\textbf{Cloud testing.} 
Most cloud testing works focus on finding bugs and vulnerabilities in the cloud service implementation itself, using stateful fuzzing~\cite{restler, auto_test,rest_checking, rest_combinatorial, rest_intelligent,restlearn} (more so with LLMs~\cite{restless, terrafault}) or differential testing~\cite{diff_test}. 
The closest efforts to ours aim to find behavioral gaps between real cloud and emulators ~\cite{icse2025}, but do not fix them in a principled manner.
Our key contribution is to close the loop: we use the discrepancies to refine our learned model itself. We take inspiration from the prior work on protocol reverse engineering~\cite{fsm_reverse, fsm_reverse_2, fsm_reverse_3} for efficient automated alignment.

\section{Conclusion} 
\label{sec:conclusion} 

Cloud DevOps is an safety-critical task, requiring DevOps programs to be thoroughly tested and validated before deployment. Directly testing against the actual cloud incurs cost and high latency, and may introduce unwanted side effects if the tests fail (e.g., risks to existing infrastructure, the need for cleanup and redeployment). Cloud emulators have gained popularity as an alterantive, which run API-level simulations locally in a lightweight manner. However, this task requires constant, error-prone manual efforts, which struggles to catch up with the cloud API surface, which is not only complex, evolving, but also differs across providers.
\sys is a new approach, using a neurosymbolic approach to learning the emulation logic from cloud documentation and automatically aligning the emulator to the cloud behavior. We show the effectiveness of \sys against existing emulators, and its generalizability across two major clouds.

\bibliographystyle{acm} 
\bibliography{citation, proposal}

\end{document}